\documentclass{kluwer}    % Specifies the document style.

\newdisplay{guess}{Conjecture}

\usepackage{graphicx}
\begin{document}                                                                                   
\begin{article}
\begin{opening}         
\title{Measuring CMB polarisation with the Planck mission} 
\author{Jacques Delabrouille}  
\runningauthor{Jacques Delabrouille}
\runningtitle{CMB Polarisation with Planck}
\institute{PCC --- Coll\`ege de France, 11 place Marcelin Berthelot, F--75231 Paris cedex 05 }
\date{April 10, 2003}

\begin{abstract}

In this paper, we discuss why and how the Planck mission, originally designed and proposed for mapping intensity fluctuations, has been revised for polarisation measurement capability as well. 
     
\end{abstract}
\end{opening}           

\section{Introduction}  

After the great success of Cosmic Microwave Background (CMB) anis\-otropies as a data goldmine for constraining cosmological models, a large fraction of the CMB experimental community now turns towards building experiments for measuring the small fraction of polarised emission in the CMB. Motivations for doing so are many, but the experimental effort quite challenging. 

Polarisation signals are much weaker than temperature anisotropies.
The amplitude of polarisation fluctuations from scalar modes is about one tenth of that of CMB anisotropies on small scales, and much lower (relatively) on large scales. The most interesting polarisation signature, that of gravity waves from inflation, is even smaller by yet an order of magnitude, possibly more. Hence, very sensitive instruments are needed to measure the polarisation signals of the CMB.

The Planck mission, to be launched by ESA in 2007, is the third generation CMB mission after COBE and WMAP.
In this paper, we review the motivations for CMB polarisation observations and the design and expected outcome of the Planck mission for polarisation measurements.

\section{CMB polarisation: Theoretical review}

The standard paradigm for the generation of CMB anis\-ot\-ropies is now well understood. Initial perturbations, generated during an inflationary epoch in the very early Universe, evolve under the effect of gravity, which causes over--densities to collapse, opposed by radiation pressure, which prevents structures from collapsing as long as radiation (mainly photons) is tightly coupled to matter. These opposing forces generate acoustic oscillations in the photon--baryon fluid prior to decoupling, giving rise to peaks and troughs in the anisotropy spatial power spectrum.

\subsection{The generation of CMB polarisation}

Temperature anisotropies on the sky are essentially generated through the combination of three effects. First, regions of the Universe intrinsically hotter at last scattering emit more and hotter photons. Second, photons coming out of overdensities are redshifted by gravitational effect, while those coming from underdensities are blueshifted. Finally, fluid motions at last scattering generate an additional redshift or blueshift by Doppler effect.

On large scales, causal processes have had no time to change much the primordial fluctuations. Hence, the CMB anisotropies reflect essentially the original perturbations (with some modifications due solely to the Integrated Sachs-Wolfe effect on large scales). On sizes smaller than the horizon, however, original perturbations have been modified by a transfert function which depends on the physics governing the acoustic oscillations. Hence, the relative height of the acoustic peaks depend on the mass/energy density content of the Universe (of baryons which are the main source of inertia, of photons which are the main source of radiation pressure, of dark matter which contributes most of the depth of the potential wells during the matter domination epoch) and on the strength of the damping term due to the expansion (and hence on the Hubble constant). Finally, the translation from physical sizes to angular sizes depends on the geometry, and hence on cosmological parameters as the curvature and the cosmological constant.

To some extent, the generation of polarisation is simpler than that of temperature anisotropies. Polarisation can be generated solely at last scattering by the dependence of the Thomson cross section upon the the polarisation states of the incoming and outgoing photons, $\sigma_T \propto |\vec{\epsilon_i}.\vec{\epsilon_o}|^2$. Hence, a quadrupole in the radiation intensity at some point at last scattering generates polarisation in a direction perpendicular to that of the maximum of the quadrupole tangentially to the last scattering surface \cite{1997NewA....2..323H}.

Contrarily to CMB temperature fluctuations which arise from a superposition of monopoles, dipoles (from Doppler effect) and quadrupoles at last scattering, CMB polarisation arises from quadrupoles alone. Hence, polarisation is an essential tool to separate between sources of anis\-otropies (monopoles, dipoles, quadrupoles), and lift some of the possible degeneracies among cosmological parameter sets as compared with constraints obtained from CMB anisotropies alone.

\subsection{Polarisation description}

\subsubsection{Stokes parameters}

The polarisation state of an incoming transverse electromagnetic wave is fully described in a given reference frame by four parameters $I$, $Q$, $U$, and $V$ defined as follows:

\begin{eqnarray}
I = \langle |E_x|^2 \rangle + \langle |E_y|^2 \rangle \\
Q = \langle |E_x|^2 \rangle - \langle |E_y|^2 \rangle \\
U = \langle E_xE_y^\dagger \rangle + \langle E_yE_x^\dagger \rangle \\
V = i(\langle E_xE_y^\dagger \rangle - \langle E_yE_x^\dagger)\rangle
\end{eqnarray}

$I$, $Q$, $U$, and $V$ are known as the Stokes parameters of the radiation. $I$ gives the total intensity, $Q$ and $U$ describe the linearly polarised part, and $V$ the circularly polarised part of the radiation.

The measured Stokes parameters depend on the reference system ($X$, $Y$).
In a frame ($X'$, $Y'$) obtained from ($X$, $Y$) by a rotation by an angle $\psi$, the Stokes parameters $I'$, $Q'$, $U'$, and $V'$ are obtained from $I$, $Q$, $U$, and $V$ by the following transformation:
$$
\left( \begin{array}{c} I' \\ Q' \\ U' \\ V' \end{array} \right) = 
\left( \begin{array}{cccc} 
1 & 0 & 0 & 0 \\
0 & \cos2\psi & \sin2\psi & 0 \\
0 & -\sin2\psi & \cos2\psi & 0 \\
0 & 0 & 0 & 1 \end{array} \right)
\left( \begin{array}{c} I \\ Q \\ U \\ V \end{array} \right)
$$
Hence, $Q$ and $U$ are defined in a frame--dependent way. At each point of the sphere, typically, directions along spherical coordinates $\theta$ and $\phi$ can be used as reference -- although other choices can be made. As a consequence, changing frames from ecliptic to galactic coordinates (and redefining correspondingly coordinates $\theta$ and $\phi$ and polarisation reference axes at each point) requires computing in each pixel of the new map a position dependent linear combination of $Q$ and $U$ in the corresponding pixel of the original map. More generally, this will be the case in any map transform which does not preserve reference axes directions. Local projections onto small square maps for instance, especially near the poles where there is a coordinate singularity, involve the same kind of difficulties. Hence, the reprojection and/or repixelisation is even more technical for polarisation maps than for temperature maps.

\subsubsection{Spherical harmonics expansion}

It has now become traditional to expand CMB temperature fluctuations on a basis of spherical harmonics on the sky. Similarly, it is desirable to expand polarisation fields. A method for doing so has been proposed by Seljak and Zaldarriaga \shortcite{1997PhRvL..78.2054S}, and independently by Kamionkowski, Kosowski and Stebbins \shortcite{1997PhRvD..55.7368K}.

The fact that $Q$ and $U$ depend on the coordinate system at each point of the sphere is a complication for a meaningful expansion. However, combinations $Q \pm iU$ (known as the Ku{\v s}{\v c}er parameters \cite{kuscer}) transform, under a rotation of coordinates by an angle $\alpha$, as $Q \pm iU \longrightarrow e^{\mp2i\alpha}(Q \pm iU)$.
Hence, $Q \pm iU$ are eigenmodes of the ``rotation by angle $\alpha$" operator on the plane, associated to eigenvalues of $e^{\mp2i\alpha}$. They are spin $\pm 2$ quantities, and can be decomposed on the sphere on the basis of spin 2 spherical harmonics $_\pm 2 Y_{\ell m}$ as:

\begin{equation}
Q(\theta, \phi) \pm iU(\theta, \phi) = \sum_{\ell,m} \, a_{ {\pm 2}, \ell m} \, _{\pm 2} Y_{\ell m}(\theta, \phi)
\end{equation}

The $E$ and $B$ polarisation fields can be defined as

\begin{equation}
E = \sum_{\ell m} a_{\ell m}^E Y_{\ell m}(\theta, \phi)
\label{eq:E}
\end{equation}
and
\begin{equation}
B = \sum_{\ell m} a_{\ell m}^B Y_{\ell m}(\theta, \phi)
\label{eq:B}
\end{equation}
where
\begin{equation}
a_{\ell m}^E = - \frac{a_{ {2}, \ell m}+a_{ {- 2}, \ell m}}{2}
\label{eq:a-E}
\end{equation}
and
\begin{equation}
a_{\ell m}^B = i \frac{a_{ {2}, \ell m}-a_{ {- 2}, \ell m}}{2}
\label{eq:a-B}
\end{equation}
The $E$ and $B$ fields are respectively a scalar and a pseudo--scalar field on the sphere (i.e. fields with respectively even and odd parity). Note that the definition of $a_{\ell m}^E$ and $a_{\ell m}^B$ given in equations \ref{eq:a-E} and \ref{eq:a-B} is just a particular choice for obtaining a scalar and pseudo--scalar field from $a_{ {2}, \ell m}$ and $a_{ {-2}, \ell m}$. Any other choice $a_{\ell m}^E \longrightarrow \alpha(\ell) a_{\ell m}^E$ would be just as good, but we stick here to the definition of Seljak and Zaldarriaga -- which is also the one used in the CMBfast software.

The intensity and polarisation of the sky is fully described by the set $I$,$E$,$B$,$V$. In turn, the spatial power spectrum of the intensity (traditionally expressed as a temperature fluctuation in units of $\mu$K, and noted $T$) and polarisation field (in the same units) of the CMB is given by:

$$
C_{\ell}^{XY} = \langle a_{\ell m}^X a_{\ell m}^{Y\dagger} \rangle =
\left( \begin{array}{cccc} 
C_{\ell}^{TT} & C_{\ell}^{TE} & 0 & 0 \\
C_{\ell}^{TE} & C_{\ell}^{EE} & 0 & 0 \\
0 & 0 & C_{\ell}^{BB} & 0 \\
0 & 0 & 0 & 0 \end{array} \right)
$$

In this spatial power spectrum, all $V$ terms vanish because no $V$--type polarisation is expected in the CMB, and all cross terms involving $B$ vanish for parity reasons. Hence, the intensity and polarisation power spectrum of the CMB is fully described, for each $\ell$, by four quantities, $C^{TT}_\ell$, $C^{TE}_\ell$, $C^{EE}_\ell$, and $C^{BB}_\ell$.

For a given cosmological model, it is possible to predict all of these terms of the matrix $C_{\ell}^{XY}$, which we refer to as the {\emph {multivariate spatial power spectrum}} of CMB temperature and polarisation.

\subsection{Science with CMB polarisation}

It is clear that polarisation and temperature anisotropies together carry more information than temperature anisotropies alone. In this section, we discuss the main scientific case for CMB polarisation measurements.

\subsubsection{Reionisation}

As polarisation is generated only at last scattering, it probes last scattering in a more direct way than anisotropies alone. In particular, polarisation provides a strong test for the ionisation history, as a significant reionisation optical depth will result in polarisation generation closer to the observer than the last scattering surface, at a later time. Large scale polarisation patterns are generated in this way, detectable as low-$\ell$ bumps in the polarisation power spectra. Recently, the WMAP team has announced a surprisingly early reionisation at a redshift of about 20 as adjusted on the best fit optical depth $\tau \simeq 0.17$ \cite{spergel}. 

If confirmed, this result also has an impact on the 
trade--off between sky coverage and sensitivity per pixel for the detection of polarisation B--modes, which may provide a case for spaceborne observations rather than ground based. As argued by Lewis, Challinor and Turok \shortcite{2002PhRvD..65b3505L}, in absence of reionisation, a deep observation of the polarisation in a small few degree by few degree patch of the sky (feasible from ground) would be more optimal to detect inflationary $B$ modes. With early reionisation, there is a large bump at very low $\ell$ which can only be measured accurately with large sky coverage, and hence from space.

\subsubsection{Cosmological parameters}

Polarisation power spectra carry also information about cosmological parameters which is partly redundant and partly complementary to that imprinted on CMB temperature anisotropies alone.

Redundancy in the information comes from the fact that temperature and polarisation anisotropies are generated from the same set of initial perturbation evolved until last scattering by the same physical processes. Hence, the pattern of peaks and troughs is connected to cosmological parameters about as unambiguously  in polarisation power spectra as in the temperature power spectrum.
Complementarity comes from the fact that the dependence of each of the spectra ($TT$, $TE$, $EE$, $BB$) on cosmological parameters suffers from different degeneracies. The joint information from all spectra hence permits to lift some of them. In addition, it provides consistency tests on the model. 

Of course, as long as errors on parameter determination 
originate essentially from instrumental noise, the accuracy of the determination of polarisation power spectra is significantly worse than that of the temperature power spectrum, and most of the cosmological information comes from the $TT$ power spectrum. The other extreme is the situation in which errors originate essentially from cosmic variance. Then, all spectra carry about as much information content, although not on the same parameters, as discussed below.

\subsubsection{Constraining inflation} \label{par:inflation}

One of the most interesting aspect of CMB polarisation is the possibility to access directly some of the inflationary parameters.

There is more and more compelling evidence that initial perturbations indeed occurred during an early phase of rapid accelerated expansion. The observation of several acoustic peaks in the CMB spatial power spectrum, the Gaussianity of the perturbations, and the spatial flatness of the Universe, all of which are strong predictions of inflation, are now well established.

Still, there are at present many inflationary models, none of which is fully satisfactory yet (in particular because of the lack of insight on the nature of the inflationary field), with too few observational means of discriminating among them. It turns out that CMB polarisation provides strong constraints on a large class of inflationary scenarios -- and can possibly permit one to measure some of the fundamental parameters of the inflationary field.

Most inflationary models indeed predict that initial perturbations comprise a contribution from tensor modes (gravity waves) in addition to scalar modes (density perturbations). Tensor and scalar modes contribute to the anisotropy spectrum on large scales. 

For a large class of inflationary models (slow-roll models), inflation is described by three main parameters \cite{liddle-lyth}

$$
\epsilon = \frac{m_{\rm pl}^2}{16\pi} \left[ \frac{V'}{V} \right]^2
$$
$$
\eta = \frac{m_{\rm pl}^2}{8\pi} \left[ \frac{V''}{V} \right]
$$
$$
r=T/S
$$
where $V(\phi)$ is a potential for the inflaton field $\phi$, and $T$ and $S$ are the amplitudes of the tensor and scalar perturbations respectively. The tensor to scalar ratio, $r = T/S$, is expected to be of about 0.1 at most and depends on the inflationary model (it is connected to the energy scale of the inflation). The scalar and tensor spectral indexes, $n_S$ and $n_T$, are obtained from $\epsilon$ and $\eta$ as
$$
n_s = 1-6\epsilon+2\eta
$$
$$
n_t = -2\epsilon
$$

Hence, for this class of model, measuring the tensor to scalar ratio $T/S$ and the spectral indices $n_S$ and $n_T$ constrains the inflationary potential $V(\phi)$ and the physics of inflation.

While $n_s$ can be measured from temperature anisotropies, the measurement of the contribution of tensor modes to the $TT$ power spectrum is limited by cosmic variance --- essentially because tensor modes contribute power only on large scales ($\ell < 100$ or so) as they decay upon entering the horizon. Small values of $T/S$, therefore, are not measurable on the $TT$ power spectrum. Even more so, the value of $n_T$ is not measurable on temperature anisotropies alone.

Similarly, tensor modes contribute a small amount of $E$ polarisation. In principle, as the cosmic variance on $E$ is smaller than on $T$, some constraints on $T/S$ and/or $n_T$ can be obtained from the large scale measurement of $C_\ell^{EE}$. However, it is much safer to detect directly the presence of $B$ modes of polarisation on large scales --- an unambiguous imprint of inflationary gravity waves. As scalar perturbations do not produce odd parity polarisation ($B$ modes), the detection of such polarisation would be a smoking gun for primordial gravity waves, permitting to constrain the physics of inflation.

There is, however, a caveat. It is quite possible that the ratio $T/S$ is small enough to escape detectability (because of foreground polarisation from lensing and foregrounds (see below), and of cosmic variance on the CMB temperature and $E$ polarisation spectra. Still, polarisation measurement seem at present to be one's best bet for constraining inflationary models --- even if the outcome is an upper limit on the energy scale of inflation.

\subsubsection{Lensing}\label{sec:lensing}

Lensing of the CMB polarisation field by large scale structure, which produces shear distorting the polarisation pattern, mixes the primordial $E$ and $B$ modes generated at last scattering. As a consequence, a contribution to $B$ modes on small scales is expected even if there is no primordial $B$. This effect permits to constrain the large scale structure power spectrum, quite interesting for constraining structure evolution through the ``dark ages" at high redshift. 

Unfortunately, this effect also makes the detection of primordial $B$ modes more difficult. 
If constraining inflationary $B$ modes is the prime target of a CMB polarisation experiment, the cosmic shear imprint on the CMB polarisation is considered as a contaminant. Prior information on the large scale structure power spectrum, combined with a measurement of the $E$ modes of polarisation, permits to predict and subtract statistically some of this lensing contribution to $C_\ell^{BB}$ \cite{2002PhRvL..89a1303K}.

\section{Experimental issues}

\subsection{Sensitivity}

One of the key issues for measuring CMB polarisation is the extreme sensitivity required, several orders of magnitude better than that required to measure CMB anisotropies.

At present, the best bolometric detectors built for sensitive balloon-borne CMB experiments are already photon noise limited. Therefore, reaching sensitivities better by two orders of magnitude (insufficient still for the ultimate polarisation experiment!) cannot be achieved by reducing detector noise. The gain in sensitivity requires about 10,000 times more detectors or 10,000 times more integration time than present-day balloon-borne or ground-based experiments. There are essentially two strategies for doing so: either build a very large camera with 10,000 detectors observing from the best possible site on ground, or multiply the number of detectors by an order of magnitude or two and benefit from the very long integration time that can be obtained on a space mission as compared to balloon experiments.

Planck detectors will be somewhat more sensitive than present best balloon-borne detectors -- by a factor of few -- essentially because of reduced background and hence reduced photon noise. In addition, it will have about ten to twenty times as many useful detectors as present balloons -- which essentially mesure anisotropies using the one or few best detectors on board. The total integration time will be about thirty times that of Boomerang and 1000 times that of Archeops.
Although not all of Planck detectors are polarisation sensitive (the final mapping of temperature anisotropies is the primary goal of the mission), Planck will have the capacity to map the full sky polarisation of the CMB well enough for meaningful constraints to be put on both $EE$ and $BB$ polarisation spectra.  

\subsection{Foregrounds}

For CMB anisotropies, there exists a window in the electromagnetic spectrum for which CMB dominates over foregrounds on a large fraction of the sky.
In addition, the near-independance of most emission laws on sky coordinates make component separation possible.

The situation is far from being as clear for polarisation. At low frequencies, Faraday rotation complicates the model of the emission. At all frequencies, the integration of polarised emission of foregrounds along the line of sight results in some averaging of the polarisation, which depends on the spatial coherence of polarisation directions. In contrast, intensities add--up as long as the emitter is optically thin, which is mostly the case for CMB foregrounds. As a result, the polarisation spatial power spectra of foregrounds can be different from the intensity spatial power spectra.

In all cases, sensitive multifrequency observations are required for assessing the situation as far as foregrounds are concerned. Current experiments provide useful measurements of foreground polarisation properties. Recently, the Archeops team has observed the dust polarisation at 350 GHz, finding 5--6\% dust polarisation on average in the galactic plane, and as much as 20\% in a few extended clouds \cite{archeops-polar}.

Hence, because foreground polarisation monitoring is an important aspect of CMB polarisation measurement, it is important to measure polarisation at several frequencies. Such multi--frequency measurements allows one to check that the detected polarisation has the expected derivative of a blackbody emission law. If necessary, it also permits to evaluate and separate foreground polarised emission from the observations. Planck will map the polarisation of the millimeter sky in six or seven frequency channels, which is not so easy to do from the ground (and depends on the polarisation properties of the atmosphere --- not completely known yet).

\subsection{Multi-detector observations}

Presently, the most sensitive operational detectors for CMB polarisation measurements are Polarisation--Sensitive Bolometers (PSB). These devices behave as a system of two polarimetric detectors, each of which is able to measure the intensity of the incoming radiation in one single linear direction.
Each such polarimeter measures ideally a signal given by
\begin{equation}
s(\alpha) = \frac{1}{2}(I + Q \cos 2\alpha + U \sin 2\alpha)
\label{eq:polarimeter}
\end{equation}
where $\alpha$ is the angle between the polarimeter orientation (i.e. the polarisation sensitivity direction) and the axes chosen for the Stokes parameters representation. For a PSB, the two polarimeters are oriented at $90^\circ$ relative angle, and hence measure signals
\begin{equation}
s_1 = \frac{1}{2}(I + Q \cos 2\alpha_1 + U \sin 2\alpha_1)
\end{equation}
and
\begin{equation}
s_2 = \frac{1}{2}(I - Q \cos 2\alpha_1 - U \sin 2\alpha_1)
\end{equation}
where $\alpha_1$ is the angle between the orientation of the first of the two polarimeters and the axes chosen for the Stokes parameters representation.
 
Whereas a single detector is sufficient -- in principle --to make a map of CMB temperature anisotropies, it is not the case for measuring polarisation, unless the same polarimeter can measure polarisation in different directions at different times. A single PSB, consisting of two orthogonal polarimeters, cannot either measure both $Q$ and $U$ in a single observation.

Whether one or several polarimeters are used in the end, measuring polarisation involves combining several measurements to separate the $I$, $Q$ and $U$ contributions. Clearly, at least three such measurements in each pixel are required to recover $I$, $Q$ and $U$ from the data. In general, a number $m$ of independent measurements permit to recover the Stokes parameters by inverting the system:

$$
\left( \begin{array}{c} s_1 \\ s_2 \\ ... \\ ... \\ s_m \end{array} \right) = 
\left( \begin{array}{ccc} 
1 & \cos 2\alpha_1 & \sin 2\alpha_1 \\
1 & \cos 2\alpha_2 & \sin 2\alpha_2 \\
.. & ... & ... \\
.. & ... & ... \\
1 & \cos 2\alpha_m & \sin 2\alpha_m  \end{array} \right)
\left( \begin{array}{c} I \\ Q \\ U \end{array} \right)
+ \left( \begin{array}{c} n_1 \\ n_2 \\ ... \\ ... \\ n_m \end{array} \right)
$$

Pure polarisation signals are obtained from differences of the readouts of two polarimeters oriented at $90^\circ$, or possibly from more complex linear combinations of more polarimeter data streams. 

In a more compact form, recovering $I$, $Q$ and $U$ at a given point $p$ from all readouts (PSB and other) amounts to inverting a system of the type
\begin{equation}
Y(p) = A S(p) + N(p)
\label{eq:polar-system}
\end{equation}
where $Y$ is the vector of all measurements at point $p$, $S$ is the vector of Stokes parameters at point $p$, $S=(I,Q,U)$, and $N$ the vector of noise for the measurements. In the linear combination, lines of matrix $A$ are ideally of the form
$A_i \propto (1,\cos 2\alpha_i,\sin 2\alpha_i)$ for a detector from a PSB, and $A_i \propto (1,0,0)$ for unpolarized detectors. 

The simplest approach to inverting the system is to use the pseudo-inverse $A^\sharp = [A^t A]^{-1} A^t$ of $A$. If however, the estimated value $\widehat A^\sharp$ of $A^\sharp$ (computed from an assumed value $\widehat A$ of $A$) is wrong, then $\widehat A^\sharp A$ is different of identity, and small off-diagonal terms induce a leakage of $I$ into polarisation signals. The same is true (and worse) when one deals with $T$, $E$, and $B$ as $T \gg E \gg B$ for CMB signals (at least in some of the useful regions of the spatial frequency range). Any inversion of the system with a matrix $W$ (using the pseudo--inverse $W = A^\sharp$, or the Generalised least square solution $W = [A^t N^{-1} A]^{-1} A^t N^{-1}$, or a Wiener inversion) suffers from the same problem of non-diagonal terms in  $\widehat WA$ when the estimate $\widehat W$ of $W$ is imperfect. This will be further discussed in the next section.

\section{The Planck design}

\subsection{The Planck mission}

The Planck mission, to be launched by ESA in 2007, has been designed primarily for the ultimate mapping of CMB temperature anisotropies, with an angular resolution of about 4.5 arcminutes, and a sensitivity of about $2 \times 10^{-6}$ per resolution element.

The Planck optics comprise a 1.5 meter  useful diameter off-axis gregorian telescope, at the focal plane of which are installed two complementary instruments. The HFI (High Frequency Instrument) is an array of bolometers cooled to 0.1 K with a dilution fridge, observing the sky in six frequency channels from 100 to 850 GHz, with polarisation sensitivity at 350, 220, 150, and possibly 100 GHz (decision pending for the last channel). The LFI is an array of radiometers observing the sky at 30, 44 and 70 GHz with polarisation sensitivity in all channels.

Planck will observe the sky from the L2 Sun-Earth Lagrange point, in a very stable thermal environment, away from sources of spurious radiation due to the Earth, the Moon and the Sun. The scanning is made along large circles around an anti--solar spin axis at a rate of 1 rpm. The spin axis follows roughly the apparent motion of the Sun, so that the full sky is covered in slightly more than 6 months.

\subsection{Principles of polarisation measurement}

All Planck detectors, both in the High Frequency (bolometer) Instrument (HFI) and in the Low--Frequency (radiometer) Instrument (LFI), are total--power detectors.

In the LFI, all detectors are polarisation sensitive. 
In the HFI, some detectors are polarisation sensitive.
The polarised detectors measure a signal given by eq. \ref{eq:polarimeter}, the other detectors measure a signal directly proportional to $I$.

\subsection{Instrumental set-up}\label{sec:inst}

\begin{figure}[tb]
\begin{center}
\includegraphics[width=7cm]{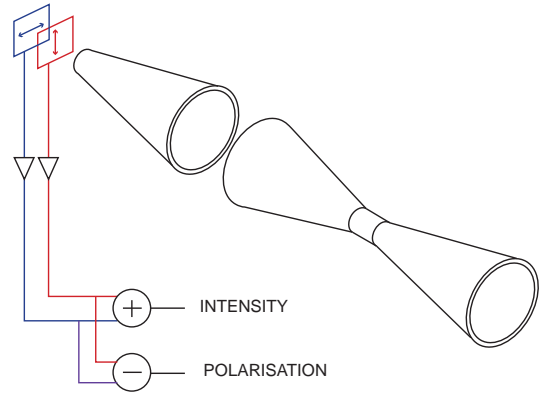}
\caption{The two detectors in a single PSB share the same optics, but have different readouts.} 
\label{fig:PSB-schema-show} 
\end{center}
\end{figure}

The Planck detector set--up takes advantage of the possibility to measure two orthogonal polarisations using the same feed horn, by splitting the orthogonal polarisations into two different detectors. For each such feed, a pair of detectors sensitive to orthogonal polarisations (polarimeters) share the same optics (filters, waveguides, corrugated horns, and telescope), but have different readouts as shown in figure \ref{fig:PSB-schema-show}.
One single such device, hence, produces two different TOI, corresponding to measurements integrated in principle over the same beam shape, but with different polarisations.

%%%%%%%%%%%%%%%%%%%%%%%%%%%%%%%%%%%%%%%%%%%%%%%%%%%%%%%%%%%%%%%%
%%%%%%%%%%%%%%%%%%%%%%%%%%%%%%%%%%%%%%%%%%%%%%%%%%%%%%%%%%%%%%%%

For the HFI, the splitting is done with Polarisation Sensitive Bolometers (PSB), in which two sensors located in the integration cavity are sensitive to orthogonal polarisations.

\begin{figure}[tb]
\begin{center}
\includegraphics[width=8cm]{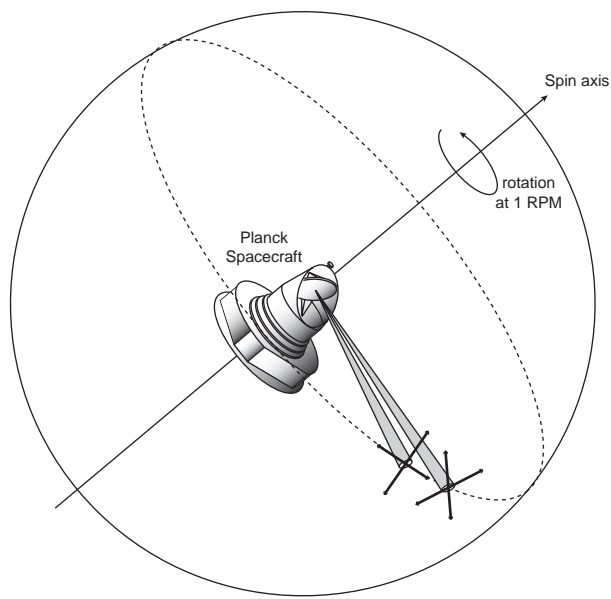}
\caption{\label{fig:planck-pol-scanning} The scanning of the polarised detectors permit to measure directly polarisation along four directions along each scan path.} 
\label{fig:planck-scanning-QU} 
\end{center}
\end{figure}

As shown by Couchot et al. \shortcite{1999A&AS..135..579C}, the measurement of polarisation using a set of polarimeters is most efficiently performed with so--called ``Optimal Configurations" (OC), using a set of $m$ polarimeters oriented at angles regularly spaced between 0 and $\pi$. A set of two feeds, with each two orthogonal polarimeters, oriented at $45^\circ$ relative angle, provides such an OC. In the Planck focal plane, most (and possibly all in a final set--up) detectors are arranged in such a way that two complementary polarimeter pairs scan the sky on the same scan path, as illustrated in figure \ref{fig:planck-scanning-QU}

%%%%%%%%%%%%%%%%%%%%%%%%%%%%%%%%%%%%%%%%%%%%%%%%%%%%%%%%%%%%%%%%
%%%%%%%%%%%%%%%%%%%%%%%%%%%%%%%%%%%%%%%%%%%%%%%%%%%%%%%%%%%%%%%%

\subsection{Aspects of data processing}

The reduction of Planck data involves several steps. The main data products to be obtained are:
\begin{itemize}
\item calibrated maps of the sky (intensity and polarisation) in all the Planck frequency channels;
\item maps of the sky (intensity and polarisation) for cosmological and astrophysical diffuse emissions (CMB, thermal galactic dust emission, synchrotron emission, ...);
\item catalogs of objects (clusters of galaxies, infrared galaxies, radio sources);
\item spatial power spectra of all components.
\end{itemize}

Level 1 of Planck data reduction consists in putting all raw data (both scientific and housekeeping) into clean and organised Time Ordered Information (TOI). Level 2 consists essentially in the map--making step which combines the level 1 outputs into calibrated maps for each frequency channels, as free as possible from noise and systematics. Level 3 is the component--separation step, and level 4 consists in delivering the data to the community at large.

We now discuss some of the polarisation--specific issues of Planck data--processing, with an emphasis on Planck HFI, although much of the discussion is appropriate for LFI data as well. 

Data processing involves three key aspects: modeling the problem, solving the problem, implementing the solution. Of these three aspects, we will now discuss to some extent the first two.

\subsubsection{Modeling the problem}
Modeling the problem consists, knowing the science issues and the way the data is measured (the instrument), to analyse the dependence function $F$ in the relation:
\begin{equation} 
{\rm data} = F({\rm instrument},{\rm science})
\end{equation} 
Both the instrument and the science can be modeled with a set of relevant parameters, some of which are known a priori, some of which are unknown. For some of the unknown parameters, statistical priors are available (e.g. uncertainties, allowed ranges, etc...).

A single sample at time $t$ from one detector in a PSB can be written as:

\begin{equation}
y(t) = h(t) \star s(t) + n(t)
\end{equation}
where $h(t)$ is the response of the detection chain (bolometer and readout electronics), $n(t)$ is noise including systematics, and $s(t)$ the signal from the sky as observed through the optics:
\begin{equation}
s(t) = \int {\rm d} \Omega {\rm d} \nu \, \left(
\widetilde I I(t) + \widetilde Q Q(t) +
\widetilde U U(t) + \widetilde V V(t) \right)
\end{equation}
In this equation, $\widetilde I$, $\widetilde Q$, $\widetilde U$, and $\widetilde V$ are the instrumental responses, assumed to be time-independent to first order. $I(t)$, $Q(t)$, $U(t)$ and $V(t)$ are the Stokes parameters of the sky in the appropriate coordinate system. The dependence of $\widetilde I$, $\widetilde Q$, $\widetilde U$, $\widetilde V$ and $I$, $Q$, $U$, $V$ on sky coordinates is not written for readability.

This model reaches a reasonable level of sophistication, although it is yet far from complete. Systematics are not specified; the response is assumed to be linear; sampling effect are not described in full detail (although the integrating window can be taken into account in the expression of $h(t)$).

It is convenient to express all quantities in the instantaneous spin--axis frame, in which case a time delay $\delta t$ can be transformed in a coordinate shift as $\delta\phi = \psi = \omega_{\rm spin} \delta t$. Then, the convolution $h(t) \star s(t)$ becomes an integral on the sky, and we have for a stable spinning:

\begin{equation}
y(t) = \left[ \int {\rm d} \psi \, h(t-\psi/\omega_{\rm spin})
\int {\rm d} \Omega \rm d \nu \sum_i
\widetilde S_i(\nu,\theta,\phi) S_i(\nu,\theta,\phi+\psi) 
\right] + n(t) 
\label{eq:full-model}
\end{equation}
Where $i$ ranges from 0 to 3, $S_0=I$, $S_1=Q$, $S_2=U$ and $S_3=V$ and Stokes parameters are expressed in a sky reference frame for which the north pole is along the spin axis.

The response to the sky signal for a stable spinning period (period for which the spin axis direction and the rotation vector magnitude do not change with time) is the polarized convolution,  in the spin-axis frame, of the instrumental beam with the sky smoothed in the $\phi$ direction by the impulse response $h$.

The connection of this equation with equation \ref{eq:polarimeter} is simple: Instead of being a known linear combination of Stokes parameters at a single point, measurements $y$ are a weighted sum of Stokes parameters in a neighborhood in space and frequency (i.e. inverse wavelength). The responses  $\widetilde S_i$ and $h(t)$ specify the elements of matrix $A$ of equation \ref{eq:polar-system}. Equation \ref{eq:polar-system} is an approximation valid under the assumption that Stokes parameters are constant in that neighborhood.

Polarimeters do not measure pure polarisation signals. The intensity response $\widetilde I$ is always larger than (or of the same order of magnitude as) $\widetilde Q$ and $\widetilde U$ for a polarimeter in a PSB, and the sky intensity is much larger than its polarisation. Neglecting the $V$ term assumed to be negligible, combining polarimeter measurements into pure polarisation signals $Q$ and/or $U$ amounts to eliminating $I$ from a set of equations as eq. \ref{eq:full-model}, in which responses 
$\widetilde I$, $\widetilde Q$, $\widetilde U$, $\widetilde V$ 
and $h$ are not so well known a priori. This can be viewed as a {\emph component separation} problem in which one of the components (the temperature signal $I$) is much stronger than the others and in which the mixing matrix is not perfectly well known.

Hence, polarisation measurements simply ``falling in the same pixel" cannot be straightforwardly combined to get polarisation by inverting the approximate model of equation \ref{eq:polar-system}, because of the risk of a significant leakage of intensity $I$ signals into polarisation signals (imperfect separation) in several ways:

\begin{itemize}
\item If beams are different (absolute difference), the inversion yields an error proportional to $I$ and to the beam difference.
\item If beams are identical, but oriented differently (case of a detector with an asymmetric beam ``coming back" to the same point with a different orientation), the inversion yields an error proportional to $I$ and to the difference of the beams in the two orientations (which has a quadrupole component, mimicking the effect of a quadrupole at last scattering).
\item If optical beams are identical and similarly oriented, but time responses $h(t)$ are different, the effective beams for the measurements are different. This yields an effect similar to the two previous cases.
\item If optical beams are identical and similarly oriented, time response effects assumed to be negligible, but pointings slightly different inside the same pixel, then the inversion yields an error proportional to the gradient of $I$ inside the pixel.
\item If the shape of the responses $\widetilde I$, $\widetilde Q$, $\widetilde U$, $\widetilde V$ as a function of frequency is different (different spectral responses), the inversion yields an error proportional to $I$ integrated in the spectral response difference.
\end{itemize}

Note that although such effects exist also for temperature measurements, they are much more critical for polarisation, the effect being typically small as compared to the magnitude of $I$. Such effects that are below sample variance and foreground--induced errors for $I$ may be above sample variance and foreground--induced errors for polarisation.

\subsubsection{Solving the problem}
Solving the problem amounts to finding the theoretical solution for estimating at best the useful parameters given the data and the priors.

In Planck data processing, we take advantage from the fact that PSBs (or LFI radiometers) comprise two polarimeters sharing the same optics. Accurate models of the optical responses and measurements of the detector performance with Archeops and Boomerang indicate that the responses $\widetilde I$ of the two polarimeters in a same PSB are very similar, minimising the impact of the above systematics in the data as far as direct differences of the TOI from the two detectors in the same horn are concerned.

Such differences are direct measurements of $Q$ and $U$ in the instantaneous spinning reference frame (the frame for which the North pole is along the spin axis). These measurements can then be reprojected on polarisation maps directly, avoiding most of the $I$ leakage problem.

This nice design feature, however, does not solve all the problems of polarisation data processing. Some aspects of the data processing are discussed in Revenu et al. \shortcite{2000A&AS..142..499R}, which investigate the suppression of stripes due to low frequency drifts, in Rosset et al. \shortcite{rosset}, which investigate the effect of imperfect beams, of pointing inaccuracies, and of response mismatches in general, in Prunet et al. \shortcite{2000MNRAS.314..348P}, which discuss a Wiener--based component separation of polarised emissions from CMB and foregrounds assuming prior knowledge of emission laws and power spectra, in Baccigalupi et al. \shortcite{bacci}, which discuss the blind separation of polarised foregrounds specifically, in Lewis et al. \shortcite{2002PhRvD..65b3505L}, which discuss the estimation of $E$ and $B$ spatial power spectra on incomplete sky maps.

\subsection{Science with Planck}

\begin{figure}[tb]
\begin{center}
\includegraphics[width=10 cm]{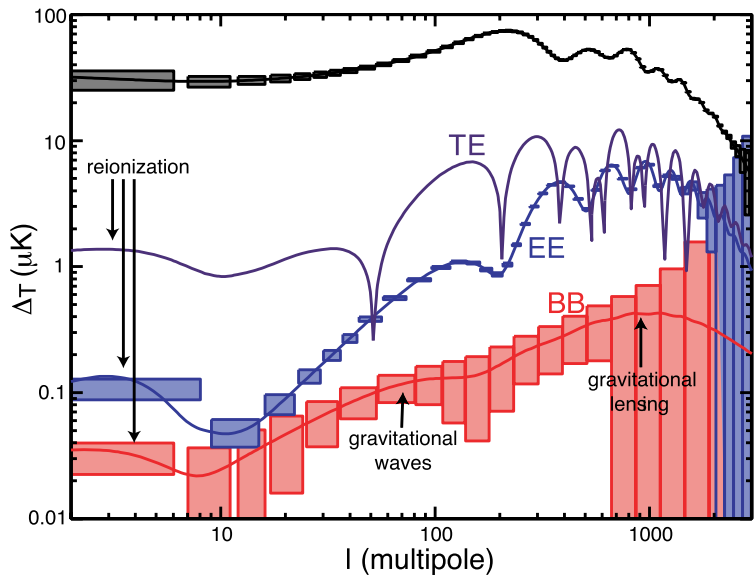}
\caption{The four non-vanishing elements of the multivariate power spectrum of CMB temperature and polarisation for a particular cosmological model, with error boxes expected from Planck observations.
} 
\label{fig:spectres-polar}
\end{center}
\end{figure}

Predictions of the accuracy on temperature and polarisation spectra with Planck are shown in figure \ref{fig:spectres-polar} (adapted from \cite{2002ARA&A..40..171H}). Whereas error boxes are large especially on $B$, meaningful constraints can be put on cosmological scenarios. Note that this figure assumes an optimistic energy scale of inflation $E_i = 2.2\times 10^{16}$ GeV, but reionization later than what announced from recent WMAP results.

%%%%%%%%%%%%%%%%%%%%%%%%%%%%%%%%%%%%%%%%%%%%%%%%%%%%%%%%%%%%%%%%
%%%%%%%%%%%%%%%%%%%%%%%%%%%%%%%%%%%%%%%%%%%%%%%%%%%%%%%%%%%%%%%%

\section{Conclusion}

Measuring CMB polarisation is the next objective of the study of the cosmic microwave background. There are several motivations for doing so. 

First, polarisation power spectra depend on cosmological parameters in a different way than the temperature power spectrum. Hence, they provide complementary information about cosmology to temperature anisotropies alone, lore and lore useful as the fundamental limitations come from cosmic variance and parameter degeneracies rather than experimental noise. In particular, early reionisation has a characteristic imprint on low $\ell$ modes of CMB polarisation and temperature-polarisation power spectra. Also, polarisation permits to separate CMB anisotropy contributions induced by Doppler shifts (which do not produce polarisation) from those intrinsic to density and tensor perturbations (which do produce polarisation).

The $B$ modes (odd parity) CMB polarisation field at low $\ell$ can be generated solely by primordial gravity waves (in inflationary scenarios). At high $\ell$, the weak shear due to structures in the Universe distorts the polarisation pattern, turning a small fraction of the $E$ type polarisation into $B$ type. Hence, measuring polarisation on all scales permits both to constrain the amount of gravity waves from inflation, as well as constrain the power spectrum
P(k) of large scale structure, if the two effects can be properly disentangled (which is possible if $T/S$ is not smaller than about 0.001).

The Planck mission will map the polarised sky in six or seven frequency channels (30, 44, 70, 150, 220, 350 and possibly 100 GHz). The sensitivity of Planck to polarisation, of the order of ten microkelvin per resolution element on average and about ten times more in highly redundant patches, will permit to put strong constrains on the polarisation power spectra of the CMB.

\theendnotes

\end{article}
\end{document}